\documentclass[11pt]{article}
\usepackage{moriond,epsfig}

\newcommand{\mincir}{\raise -2.truept\hbox{\rlap{\hbox{$\sim$}}\raise5.truept
\hbox{$<$}\ }}
\newcommand{\magcir}{\raise -2.truept\hbox{\rlap{\hbox{$\sim$}}\raise5.truept
\hbox{$>$}\ }}
\newcommand{\fl}{\,{\rm erg\,s^{-1}cm^{-2}}}
\newcommand{\hm}{\,h^{-1}{\rm Mpc}}
\newcommand{\vel}{\,{\rm km\,s^{-1}}}

\def\be{\begin{equation}}
\def\ee{\end{equation}}
\def\bea{\begin{eqnarray}}
\def\eea{\end{eqnarray}}

\begin{document}
\vspace*{4cm} \title{Summary: Clusters of Galaxies and the High Redshift
Universe\\Observed in $X$--Rays}

\author{ S. Borgani }

\address{INFN - Sezione di Trieste, c/o Dipartimento di Astronomia
dell'Universit\`a\\ via Tiepolo 11, I-34131 Trieste, Italy}

\maketitle\abstracts{This Meeting featured the recent advancements in
our understanding of galaxy clusters and the distant Universe,
achieved by the past and new generation of $X$--ray satellites. I
summarize here the main themes that have been discussed: {\em (a)}
Clusters of galaxies as probes of cosmological models; {\em (b)} The
physics of cosmic baryons trapped within the potential wells of galaxy
clusters; {\em (c)} The origin of the cosmic $X$--ray background and
the nature of the contributing sources.}

\section{Introduction}
$X$--ray extragalactic astronomy is experiencing a sort of
revolution. The unprecedented observational capabilities, which are
offered by the Chandra and the XMM--Newton satellites, are triggering
a significant change of perspective in our understanding of the
processes of formation and evolution of cosmic structures. At the same
time, $X$--ray observations are complemented and integrated by data at
other wavelengths, from radio, to optical, to EUV and
$\gamma$. Undoubtedly, this Conference has been organized in the right
moment (and in the right place!) to gather people working on frontier
observations and modelling of galaxy clusters and on high--redshift
$X$--ray sources.

As for galaxy clusters, we are approaching the full exploitation of
the data flow provided during the last decade by the {\tt ROSAT}, ASCA
and Beppo--SAX satellites. At the same time, the $X$--ray observations
realized during the last year or so with the Chandra and XMM--Newton
satellites are drastically improving our knowledge of the physics of
the intra--cluster medium (ICM). The unprecedented spatial resolution
of the Chandra satellite reveals a variety of small scale details
in the distribution of the baryonic component of clusters, which are
witnessing the complex dynamical and physical processes taking place
in cluster central regions. At the same time, the good spectral
resolution of XMM--Newton, joined with a good spatial resolution,
allows an order-of-magnitude refinement in the mapping of the
temperature and cooling structure of the ICM, and of the pattern of
its metal enrichment. There is no doubt that such a detailed
description of the ICM physics is challenging the theorist's view of
clusters as arising from semi--analytical and numerical
modelling. Furthermore, the possibility of using clusters as
high--precision probes of cosmological models relies on an accurate
understanding of the connection between their internal properties,
which are affected by gas physics, and the cosmic evolution of the
global structure of the Universe, which is mainly driven by gravity.

Besides the detailed investigation of nearby objects, deep $X$--ray
pointings are now peering deep into the distant Universe and are
unveiling the evolution of different population of cosmic
structures. A handful clusters at $z>1$ have been secured, mostly by
{\tt ROSAT} data, and there is little doubt that their number will
significantly increase in the next future. Observations of such
extreme clusters with satellites of the last generation are tracing
the epoch at which they shape out from the cosmic web and make baryons
switching on in the $X$--ray light.

Finally, deep pointed observations are nowadays providing the solution
to the long-standing debate about the nature of the $X$--ray
background (XRB). The spatial resolution of Chandra is now resolving a
large fraction of the XRB into discrete sources, up to energies of
about 7 keV. At the same time, XMM deep fields are extending this
study to higher energies, $\simeq 12$ keV, therefore recovering the
very hard sources which are missed by the softer energy coverage of
Chandra.

\section{Cosmology with $X$--ray galaxy clusters}
Using $X$--ray galaxy clusters for cosmological purposes has two main
advantages with respect to clusters selected in the optical
band. First of all, $X$--ray emissivity is proportional to the square
of the local gas density, while optical light has, to a good
approximation, only a linear dependence on the galaxy density. This
makes clusters looking sharper in the $X$--ray than in the optical
sky, with the obvious consequence of reducing projection contamination
and allowing for a better definition of the sample selection
function\cite{Donahue}. Furthermore, the $X$--ray luminosity, $L_X$,
is better correlated to the total collapsed cluster mass, both from a
theoretical and a phenomenological point of view, than the optical
(i.e., Abell--like) richness. Press-Schechter--inspired approaches and
N--body simulations allow cosmological models to predict statistical
properties (e.g., number density and correlation function) for the
cluster population of a given mass. Therefore, $X$--ray clusters have
the obvious advantage of being selected according to an observable
quantity, $L_X$, which is well correlated with the theoretically
predicted mass.

\subsection{Large--scale structure}
The {\tt ROSAT} All Sky Survey (RASS) has been a sort of gold mine
over the last decade for the extraction of large samples of galaxy
clusters. The most recent of these samples, the REFLEX and the NORAS
surveys, contain $\sim 1000$ clusters, down to a flux of a few$\times
10^{-12}\fl$ ([0.1-2.4] keV energy band), and allow to trace the
large--scale structure (LSS) of the Universe out to $\sim 10^3$ Mpc,
through the estimate of the two--point correlation function and the
power spectrum \cite{Schuecker,Retzlaff}. Although a firm detection of
a large--scale turn-over of the power spectrum has still to be
confirmed, nonetheless there is a clear sign that it stops
continuously increasing toward small wavenumbers, thus witnessing the
``end of greatness'' in the cosmic LSS. Taking advantage of the
$X$--ray selection, it is then possible to compute the biasing factor
for these clusters once a cosmological framework is assumed
\cite{BorganiGuzzo}. As a result, it is shown that the clustering of
$X$--ray clusters is consistent (surprise, surprise!)  with a
low--density CDM model, with $\Omega_m\simeq 0.3$, flatness restored
by cosmological constant, and $h\simeq 0.7$
\footnote{As usual, I define here the Hubble parameter as
$h=H_0/100\vel {\rm Mpc}^{-1}$}.

An intrinsic limitation of available wide--area samples of $X$--ray
clusters lies in their limited depth (median redshift $z\simeq 0.1$),
which is definitely too shallow to probe cosmology through the
redshift evolution of the LSS. Further squeezing the RASS to reach
fainter fluxes shouldn't help much in this respect. A real step forward
should require, instead, a survey--dedicated $X$--ray satellite, which
would be able to search for clusters over a large ($\sim 10^4$
sq. deg.) contiguous area, while reaching fluxes two orders of
magnitude fainter than the RASS. A satellite with these
characteristics is surely within the reach of available technology and
there are good chances for it to become reality in the next future.

\subsection{Evolution of the cluster population}
A complementary approach to constrain cosmological models with galaxy
clusters is represented by the evolution of their number density or,
formally speaking, the evolution of the cluster mass function. After
the pioneering work of the Einstein Medium Sensitivity Survey (EMSS),
several independent $X$--ray samples have been constructed over the
last 5 years, all based on deep {\tt ROSAT} exposures (e.g., RDCS, 160
sq.deg. CfA, SHARC, WARPS, NEP, BMW), each containing $\sim 50$--100
clusters\cite{Schuecker,Rosati,Adami}. Thanks to their fairly large redshift
baseline ($z\mincir 1.3$), they provide nowadays the most effective
probe for the evolution of the cluster population. Since cluster
evolution depends on the growth of the cosmic density perturbations over
$\sim 10 \hm$, it constrains the r.m.s. density perturbation over this
scale, $\sigma_8$ \footnote{I use here the standard definition of
$\sigma_8$ as the r.m.s. density fluctuation within as top--hat sphere
of $8\hm$ radius}, and the cosmic density parameter, $\Omega_m$, while
being much less sensitive to the value of the cosmological constant
\cite{Blanchard,Lukash,Diego,Douspis}.

Analyses of the redshift dependence of the $X$--ray cluster luminosity
function for these samples, based on the Press--Schechter (PS)
approach or on refinements of it, are providing constraints on
$\sigma_8$ and $\Omega_m$. Although some debate is still present, I
think it is fair to conclude that values of the density parameter
$\Omega_m\magcir 0.5$--0.6 are disfavored by these analyses at the
$3\sigma$ level. Still, such results are not providing a precision
determination (i.e., with $\mincir 10\%$ accuracy) of cosmological
parameters, the main source of error being represented by the limited
size of the samples and the uncertainties which are anyway present in
the relation between $L_X$ and cluster mass.  In order to overcome
these sources of uncertainties, one should look for a better
calibration of this relation, for instance through weak lensing
determinations of cluster masses \cite{Bridle}.

As a further possibility, one can resort to the $X$--ray temperature
of the ICM, $T_X$, which is better connected than $L_X$ to cluster
mass, once hydrostatic equilibrium is assumed. Follow up observations
to determine $T_X$ for samples of nearby clusters have been already
realized, while extensions to include a fair number of distant
($z\magcir 0.5$) clusters are still underway. However, precise
estimates of $T_X$ at high redshift are observationally quite
demanding. Furthermore, the relation between $T_X$ and collapsed mass
is also prone to some degree of uncertainty, connected to complex
cluster dynamics, temperature gradients, cooling--flow structures,
etc. Finally, analytical methods also introduce some degree of
approximation in the theoretical description of the distribution of
cluster virial masses, which propagates into errors in the
determination of cosmological parameters.

What should we do, then, to improve the analysis of the cluster
evolution and make it competitive with other high--precision
cosmological tests, such as the spectrum of CMB fluctuations?

\subsection{High--precision cosmology with galaxy clusters?}
There is no doubt that statistics of distant clusters will be no
longer an issue in years to come: XMM--Newton and Chandra data will
allow to enlarge {\tt ROSAT}--based samples of high--$z$ clusters by
at least one order of magnitude
\cite{Valtchanov,Bartlett}. Furthermore, other methods to identify
distant clusters, based e.g. on the Sunyaev--Zeldovich effect (see
below), will push to higher $z$ the mapping of the cluster population.
At the same time, simulation campaigns aimed at describing the LSS
evolution over volumes comparable with the whole size of the
observable Universe, are providing accurate calibrations of PS--like
methods and, therefore, allows to keep under control the theoretical
uncertainties in the description of the DM halo mass function
\cite{Evrard1}.

Therefore, the possibility for the high--$z$ evolution of the cluster
population to be promoted as a precision test for cosmology will
ultimately rely on our ability to connect to each other theoretical
and observational clusters and, ultimately, to answer to the following
question.

\section{What is a cluster of galaxies?}
\subsection{Clusters with SZ}
One session of this Meeting has been devoted to cluster studies
through the Sunyaev--Zeldovich (SZ) effect \cite{Birkinshaw}. Although
this subject is far (in terms of wavelength) from $X$--ray studies,
nevertheless the two approaches nicely complement each other in
several respects. The SZ signal has been now detected at a high
significance for several tens of already known clusters out to $z\sim
0.8$, while hints of serendipitous cluster
detections have been also reported. The SZ effect, which is caused by the inverse Compton
scattering of ICM electrons onto CMB photons, manifests itself as a
distortion of the Planckian CMB spectrum in the direction of galaxy
clusters, with a decrease of the effective CMB temperature in the
Rayleigh--Jeans region. Due to its nature, the SZ effect is proportional to
the ICM pressure, integrated along the line-of-sight. For this very
reason, it has two fundamental characteristics, which make it a powerful
tool for clusters studies: {\bf (1)} The SZ signal scales linearly with
the electron number density, $n_e$, and {\bf (2)} it is essentially
independent of redshift.

The most important applications of the SZ effect, which have been
discussed at this Meeting, are the following.
\begin{itemize}
\item Determination of $H_0$ via the apparent--size distance, taking
advantage of the different dependencies of the $X$--ray and SZ fluxes
on $n_e$ \cite{Kneissl}. Although the accuracy of this method to
determine the cosmological distance scale relies on assumptions on the
cluster geometry, this uncertainty can be reduced by averaging over a
sufficiently large ensemble of galaxy clusters. First determinations
of $H_0$ from SZ provided values which were quite lower than those
from other distance indicators. Thanks to the much improved accuracy
of the analysis and quality of the data, current estimates range now
in the interval $H_0\simeq 55$--75 km s$^{-1}$Mpc$^{-1}$, thus quite
consistent with estimates from other distance indicators.
\item Determination of the gas fraction, $f_{\rm gas}$. Since the SZ
signal is less prone than $X$--ray emissivity to gas clumping, it
should provide a more robust determination of the mass fraction
contributed by hot diffuse baryons \cite{Birkinshaw}. Reported results
indicate values in the range $f_{gas}=0.05$--0.10$\,h^{-1}$, thus
consistent with $X$--ray determinations, for reasonable values of
$H_0$.  Also, the weaker dependence on $n_e$ makes the SZ signal
suitable to trace the ICM properties out to larger radii than
accessible by $X$--ray observations. Taking advantage of this
property, SZ observations could allow to study the ICM over the whole
cluster virial region, thus tracing the accretion pattern of the gas
and the gravitational shocks occurring at the cluster outskirts.
\item Detecting and mapping distant clusters, thanks to the
redshift--independence of the SZ signal
\cite{Pointecouteau,Grainger}. In principle, this feature could allow
the selection of clusters according to a genuine mass--limit criterion
\cite{Buttery}. Several ground--based SZ surveys over reasonably large
area are currently planned, which could lead to the serendipitous
detection of very distant, $z>1$, clusters. Furthermore, the Planck
satellite could provide in about five years an all-sky SZ survey.
Although this survey will be performed with worse sensitivity and
spatial resolution than reachable by ground--based observations, it
will allow accurate statistical studies from a sample containing
several thousands SZ sources.
\item Cluster peculiar velocities through the kinematic SZ effect
\cite{Birkinshaw}. The bulk motion of ICM electrons, caused by the
peculiar velocity of clusters in the CMB reference frame, generates a
distinct distortion of the Planckian spectrum, which adds to the
thermal SZ one. The amplitude of this kinematic SZ effect depends on
the cluster peculiar velocity and, therefore, could be used to map
cosmic flows through sub-mm observations. However, the size of the
kinematic SZ distortion is so small to place only weak constraints on
the peculiar velocity of individual clusters, while it would be more
effective for statistical characterization of cosmic velocity fields.
For instance, the all--sky SZ survey expected from the Planck
satellite, should provide typical errors in individual cluster
velocities of several hundreds $\vel$, while it should be able to put
significant constraints on bulk motions over large scales, $\sim
100\hm$, and even to trace its cosmic evolution out to $z\simeq 0.5$.
\end{itemize}

\subsection{The Chandra view}
The arcsec resolution achievable with the Chandra satellite is
revealing a variety of small--scale structures in the ICM, which are
witnessing the presence of complex physical processes
\cite{Forman}. Features in the gas distribution have been detected for
several clusters, although spatially resolved measurements of gas
temperature, density and pressure shows that they don't have a common
origin. In some cases (e.g., A665) complex patterns in temperature and
gas distributions clearly suggest the presence of gas shocked by an
ongoing merging. In other cases (e.g., A3667), sharp features in the
$X$--ray emissivity correspond to jumps in temperature and gas
density, which balance each other to give a continuous pressure
variation \cite{Markevitch}. Rather than to shocked gas, these
features correspond to fronts of cold gas moving within the
cluster. These fronts are associated to the presence of magnetic
fields, whose field lines run parallel to them and inhibit gas
mixing during subsonic merging.

Detailed investigations of the $X$--ray emission from the two dominant
elliptical galaxies in the Coma cluster show that it is originated
from high--density gas at a temperature $T_X\sim 1$--1.5 keV, with an
estimated cooling time $t_{cool}\sim 10^8$ yr \cite{Vikhlinin}. This
time scale turns out to be somewhat smaller than that over which
supernova (SN) explosions from the stellar population provide energy
feedback to the diffuse gas, $t_{SN}\sim 10^9$ yr. If these
order-of-magnitude estimates will be confirmed by accurate
computations, then the question would arise as to what mechanism
(e.g., suppression of heat conduction) would prevent the diffuse gas,
surrounding these galaxies, to cool down and disappear from the hot
phase. As I will discuss in the following, when reviewing results from
XMM, the physics of gas cooling is actually one of the most intriguing
open problems, and is severely challenging our understanding of the
ICM.

At a first glance, the complexities revealed by Chandra observations
would lead to conclude that real galaxy clusters are indeed much
different from the spherical and dynamically relaxed Press--Schechter
clusters, that theorists have in mind. If true, one could then wonder
whether clusters can indeed be used as cosmological probes: do they
represent fair reservoirs of the cosmic baryons? is their dynamics
known to sufficient precision to allow reliable determinations of the
collapsed mass and of the density profiles? Although real clusters
deviate from the ideal picture on small scales, there are good reasons
to consider them as fairly well behaved structures when looking at
their global properties. Indeed, Chandra data for several nearby
clusters show that: {\em (a)} baryon fraction profiles flatten already
at $\sim 0.1R_{vir}$ (cf. also ref. \cite{Sadat}); {\em (b)} DM
profiles, reconstructed from hydrostatic equilibrium equation, are
consistent with NFW profiles and, when available, also with profiles
from weak--lensing mass reconstruction \cite{Allen}. This confirms the
standard picture that clusters are fair reservoirs of cosmic baryons,
with its gas content being in equilibrium within gravitational
potential wells created by hierarchical gravitational collapse.

Besides high--resolution details of nearby clusters, the Chandra
satellite is also proving to be a powerful instrument to detect ICM
emission from high--$z$ clusters. This is demonstrated by the
observations of distant clusters reported at this Conference
\cite{Canizares,Cagnoni,Hughes}, from $z\sim 0.5$ out to the very
distant clusters at $z\sim 1.3$ detected in the Lynx field
\cite{Stanford}, and the $z\sim 1.8$ extended $X$--ray emission
identified in correspondence of a radio galaxy \cite{Fabclus}. Such
studies allow to map the structure of the ICM soon after it was
assembled from DM gravitational collapse, thus further pushing back in
cosmic time our knowledge of the evolution of the intra--cluster gas.

\subsection{The XMM--Newton view}
Due to its technical characteristics, XMM--Newton is providing
information which are complementary to those derived from Chandra data:
the larger collecting area of XMM mirrors, especially at high energy
($\magcir 5$ keV), and the better energy resolution is coupled with an
adequate spatial resolution (PSF of about 5 arcsec on axis), so as to
allow spatially and spectroscopically resolved observations of the ICM
\cite{Arnaud}. Although its instrumental background turns out to be
quite larger than expected from pre-launch calibrations, there is no
doubt that this satellite is providing unprecedented insights into the
ICM physics \cite{Nevalainen,Sauvageot,Reiprich,Bourdin,Belsole}. The
observation of the Coma cluster shown at this Meeting is a spectacular
example of such capabilities \cite{Briel}: the combination of temperature and
emission maps shows two subgroups undergoing merging along a
filamentary structure, with evidences of tidal gas stripping and bulk
shocks, much like seen in hydrodynamical simulations of hierarchical
cosmic structure formation.  Available data on the gas temperature
profiles for a few nearby clusters show that they are generally flat
out to $\sim 0.5R_{vir}$, with a drop, detected in some cases in the
central regions and associated to gas cooling
\cite{Arnaud,Pratt}. Quite remarkably, the reconstruction of the mass
density profile shows agreement with the NFW profile, at least in the
outer cluster regions. This reinforces the picture of clusters as
structures which are well behaved and dynamically relaxed on large
scales, while showing significant complexities on small scales.

Besides the determination of the gas temperature, the XMM--Newton
satellite also represents a very well suited instrument to measure the
amount and distribution of metals in the ICM and, therefore, to
reconstruct the history of its enrichment from the past star formation
within cluster galaxies \cite{Mushotzky}. The XMM energy coverage
([0.2-12] keV) is large enough to encompass atomic transitions for several
elements (e.g., $C,~Ne,~Si,~S,~Fe$). As data will be accumulating, the
possibility to determine relative and absolute abundances, along with
their spatial distribution, will allow to constrain the
contribution of supernovae to ICM metal enrichment \cite{Portinari}
and, therefore, to infer the total energy injected into the diffuse
gas from SN explosions. At present, XMM is confirming the picture of a
polluted ICM with $Z\simeq (0.3\pm 0.1)Z_\odot$ and no redshift
evolution out to $z\sim 0.5$. As for abundance gradients, they are
found to be significant only in a few cases, seemingly at variance
with respect to results from ASCA and Beppo--SAX, which instead show
evidence of gradients in most cases. As the potentiality of XMM will
be fully exploited in the next future, it will provide $Fe$ abundance
out to $z\sim 1$ and trace other elements to lower redshifts
\cite{Mushotzky}. Such potentialities are clearly demonstrated by the
XMM observation of M87 in the Virgo cluster \cite{Matsushita}, which
allowed to determine abundance gradients for six different elements.

The physics of gas cooling merit a special mention among the fields
where XMM is bringing a major contribution \cite{Fabian}. Gas in the
central part of clusters has high enough density to cool down and drop
out of the hot diffuse phase over a time scale, $t_{cool}$, shorter
than the dynamical time scale. According to the standard cooling flow
model, one should observe at each radius a superposition of gas at
different, even very low, temperatures (multiphase model), flowing
inside from outer cluster regions. If gas is left free to cool, then
$t_{cool}$ is short enough to allow a large fraction ($\magcir 50\%$)
of the whole ICM to pass to the cold phase. This is at variance with
respect to observational evidences, which indicates that only $\sim
10\%$ of cluster baryons are locked into a cold phase, mostly
contributed by stars.  The question then arises as to how one can
prevent gas from over--cooling.  In principle both SN explosions and
AGN activity could provide energy feedback to heat back the gas to the
diffuse phase.  However, although some evidence is emerging for
association of cooling gas with star--forming regions \cite{McNamara},
no evidence has been found to date for the presence of central heating
in correspondence of cooling regions. Results from XMM observations
are now adding a further complication to this puzzle: spatially
resolved spectroscopy of central regions of cooling--flow clusters are
showing no evidence for line emission associated to gas at $T_X<2$
keV. Where does the cold gas end up? Is it heated back? If this is the
case, then a large power would be needed, while we don't see any
signature of it. Although such findings could represent a serious
challenge for the multiphase model, still a good fit to measured
spectra can be provided by a model with two gas phases, one at the
virial temperature, $T_{vir}$, and the other at $\sim 1/3\, T_{vir}$
\cite{Fabian}. A proposed solution for the lack of low--$T$ metal
lines is based on assuming a bimodal distribution of ICM metals, with
small lumps of high--$Z$ gas surrounded by $\sim 90\%$ of very
low--$Z$ gas. This model could have the virtue of behaving like $\sim
0.3Z_\odot$ gas for $T_X> 3$ keV and as metal poor gas at lower
temperatures, thus justifying the lack of emission lines from
low--$T_X$ gas. Of course one has then to motivate the presence of such
metal nuggets. There is no doubt that the accumulation of more XMM
data will soon clarify once for all whether ICM cooling has to be
associated with a multiphase flow or a different picture needs to be
elaborated.

\subsection{The non--thermal ICM}
Sharp features in the gas distribution as observed in the $X$--ray
band are not the only expected consequence of merging
events. Actually, mergers represent among the most energetic cosmic
phenomena \cite{Sarazin}: merging structures collide with a velocity
of $\sim 1000\vel$ with an involved gravitational energy as large as
$\sim 10^{64}$ ergs. Although the major part of this energy goes into
thermal heating of the ICM, thus boosting its $X$--ray luminosity, a
small fraction of it can be converted into the acceleration of
relativistic electrons. Depending on their energy, a significant
fraction of such electrons is retained within the cluster regions over
a time scale comparable to the Hubble time. The resulting emissivity
from these electrons, by inverse--Compton (IC) scattering with CMB
photons, dominates over the thermal bremsstrahlung emission in the EUV
region of the spectrum, where most of the energy is stored, and in the
hard $X$--ray band ($\magcir 20$ keV).

Claims for detection of excess EUV emission by the EUVE satellite have
been reported for six galaxy clusters \cite{Durret}. The currently
favored explanation for this emission is the IC scattering of
electrons with a Lorentz factor $\gamma\sim 300$ (corresponding to
energies of about 150 MeV). Similarly to the SZ effect, this emission
is spatially more extended than the bremsstrahlung emission, thanks to
its linear dependence on the local electron number density.  More
energetic electrons, with $\gamma\sim 10^4$, are expected to produce
by the same mechanisms a hard $X$--ray (HXR) tail in excess with
respect to thermal bremsstrahlung, as well as synchrotron emission, in
case a strong enough intra--cluster magnetic field is
present. Evidences for HXR excess have been presented for the Coma
cluster and A2256 \cite{Fusco}. An alternative explanation for this
HXR excess is based on bremsstrahlung emission from high--energy
non--thermal electrons, whose origin would however be quite
unclear. The spatial localization of this emission is still poorly
determined, due to the coarse angular resolutions of the detectors
used to date. With the improvement of the angular resolution (e.g.,
with the INTEGRAL satellite), it could be possible to see whether the
HXR emission is localized in the radio emitting regions, which are
associated with the shock fronts where electrons are accelerated.

Radio emission represents a further manifestation of non--thermal ICM
behavior \cite{Sarazin}. Diffuse radio halos have been detected in
several tens of clusters and have no obvious associations with member
galaxies. These sources are commonly classified as radio halos and
relics, depending on whether they are observed in projection near the
cluster center or in its periphery. In all the known cases, such
structures are found in clusters with evidences of recent
mergers. Therefore, the widely accepted explanation for the radio
emission is based on synchrotron emission from relativistic electrons,
that are accelerated within the merger shocks and spiralize along the
lines of the intra--cluster magnetic field. This picture is also
supported by magneto--hydrodynamical simulations of a blob of radio
plasma passing through a cluster merger shock wave, and producing
radio emitting regions with morphologies resembling the observed ones
\cite{Dolag}.

In some cases, radio activity is also connected with features in the
pattern of $X$--emission \cite{Sarazin,Clarke,Govoni}. A clear example
has been recently provided by the Chandra observation of Abell 2052,
where the radio emission surrounding the central cD corresponds to
holes in the $X$--ray brightness, with shells of $X$--ray bright
features surrounding the radio--emitting region. Since the gas
pressure is observed to be continuous across such $X$--ray
discontinuities, they do not correspond to ongoing shocks. Instead, they
are likely to arise from ICM regions where gas is compressed by the
non--thermal pressure associated to radio lobes. If this additional
pressure term is a non--negligible fraction of the thermal one, then
one may wonder whether the assumption of hydrostatic equilibrium
provides a correct description of the cluster internal dynamics. Since
these features are localized around radio galaxies, instead of being
ubiquitous, one expects non--thermal support not to be dominant.

\subsection{The theorist's view}
The amount of information on the cluster physical properties, that I
have discussed so far, still deserves an adequate interpretative
framework. In this context, numerical simulations are undoubtedly
valuable instruments to describe the gravitational dynamics of DM and
basic properties of the gas physics \cite{Evrard,Schindler}. N--body
experiments including hydrodynamics are shading light on the detailed
structure of DM halos, on the effect of merging on global cluster
dynamics, on the connection between their morphology and the
surrounding environment \cite{Plionis}, and on cooling structure of
the gas \cite{Schindler}. In most cases, the treatment of the
simulated gas is such that it reacts only to the effect of
gravitational processes, like adiabatic compression and accretion
shocks. However, a variety of observational evidences demonstrates
that other physical processes should play an important role. Thanks to
the availability of specialized codes running on massive parallel
supercomputers, mass and dynamical resolution achievable nowadays in
cluster simulations seem accurate enough to warrant numerically
convergent results \cite{Tormen}. The real challenge, instead, is
represented by the inclusion of a believable treatment of more complex
physical processes, like the gas cooling and its interplay with
processes of star formation and galaxy evolution within cluster
galaxies \cite{Borgani}, or the effect of magnetic fields
\cite{Dolag,Ensslin}.

Gravity in itself does not introduce characteristic scales. Therefore,
if gravity only acts on the gas, then the ICM should behave in a
self--similar fashion. In fact, this prediction is at variance with
respect to observations. Scaling relations among global observable
quantities of the ICM, like $X$--ray luminosity, temperature, gas mass
and entropy, violate self--similarity: the gas distribution within
galaxy groups and poor clusters ($T\mincir 2$ keV) is relatively
shallower, with suppressed density and entropy excess in central
regions, with respect to hotter ($T\magcir 3$ keV) systems. The common
interpretation is that the gas should have been heated by some
non--gravitational process before the cluster collapse. The amount of
this extra energy introduces a characteristic scale into the problem
and, therefore, breaks self--similarity. Numerical simulations
including pre--heating, in the form of pre--collapse gas entropy
floor, show that an extra--energy of about 1 keV per particle is
required to reproduce observational results on the entropy excess and
the shape and evolution of the $L_X$--$T$ relation
\cite{Borgani,Evrard}. Furthermore, if this pre--heating affected the
high--redshift ($z\sim 3$) intergalactic medium, it should have left
its imprint also on the $X$--ray emission from large--scale
filamentary structures \cite{Voit}: the decreased gas density in
filaments would correspond to a suppressed emissivity and, therefore,
to a reduced contribution to the soft $X$--ray background (XRB; see
the discussion on the XRB here below).

First attempts to include the effect of SN feedback in cluster
simulations \cite{Borgani} indicate that they can hardly provide the
correct amount of heating energy, thus requiring a further
contribution from other mechanisms, such as AGNs. However, my
impression is that our current understanding of the relative role
played by type Ia and II SN and their efficiency in dumping energy
into the hot diffuse medium is not yet accurate enough to draw firm
conclusions. The improved accuracy in the determination of ICM
chemical abundances will soon provide useful insights into our
understanding of the effect of star formation on the ICM
physics. Finally, whatever the source of ICM pre--heating is, it
should act on spatial and temporal scales tuned so as to stop the
cooling runaway and to allow only a small fraction ($\sim 10\%$) of
the diffuse gas to cool down into a collisionless phase. In this
sense, gas cooling and non--gravitational heating are inextricably
linked aspects which should be accounted for in a self--consistent
way.

\section{Resolving the $X$--ray background}
Besides galaxy clusters, a session of this Conference has been devoted
to discussing the $X$--ray background (XRB) in the light of recent
Chandra \cite{Tozzi} and XMM--Newton \cite{Hasinger}
observations. Taking advantage of the exquisite angular resolution of
the Chandra satellite, independent groups are actively working on the
analysis of long--exposure pointings to resolve the contribution of
discrete sources to the soft and the hard XRB. In fact, the results
reported from the one--million seconds ACIS--S observation of the
Chandra Deep Field South (CDFS) clearly demonstrate that the most part
of the XRB is contributed by discrete sources \cite{Tozzi}. At the
flux--limit, $4\times 10^{-16}\fl$, reached in the hard ([2--7] keV)
band, 65--$95\%$ of the background is resolved, the exact fraction
depending on the assumed level for the observed
background. Furthermore, number counts show evidence of flattening at
faint fluxes, $\mincir 10^{-14} \fl$, thus demonstrating that we are
actually identifying most of the discrete sources making up the hard
XRB. Although the comparison among different fields observed with
Chandra deep pointings shows a fairly good agreement, differences have
been detected in some cases which can hardly be explained by
Poissonian fluctuations. The likely explanation should lie in the
intrinsic clustering of Chandra sources, which enhances the
field-to-field scatter of the flux number counts.

Another long--standing problem from pre--Chandra XRB observations
concerned the so--called ``spectral paradox'', that is the difference
between the hard profile of the XRB spectrum and the relatively softer
spectrum of the detected sources. Spectral analysis of the sources,
identified now by Chandra observations, shows that they become
progressively harder at fainter fluxes. Therefore, the spectral
paradox is solved by a faint population of hard sources, contributed
by $z\mincir 1$ absorbed AGNs. Such results on the hard XRB are
reinforced and extended to higher energies by the results of deep
XMM--Newton observations, like that of the Lockman hole
\cite{Hasinger}. Despite the lower angular resolution, the better
sensitivity of this satellites to harder photons provides the deepest
determination of number counts in the [5--10] keV band and resolves
about 60\% of the XRB at these energies. The identification of the XMM
sources, both in the Lockman hole \cite{Hasinger} and in the
Groth--Westphal strip \cite{Miyaji}, confirms the picture that the hard XRB
at these energies is mainly contributed by intrinsically
absorbed AGNs.

As for the soft ([0.5--2] keV) XRB, it is virtually completely
resolved ($\simeq 80$--$95\%$) by Chandra deep exposures. As already
mentioned in Section 3.5, this result provides a non--trivial
constraint on the diffuse emission from the warm diffuse gas
permeating the large--scale cosmic web \cite{Voit}. Much like for the
ICM, the suppressed $X$--ray emission is consistent with the picture
of non--gravitational heating placing the gas on a higher adiabat, and
preventing it from reaching high densities within large--scale
filamentary structures.  Finally, the knowledge of the flux and
redshift distribution of $X$--ray (and UV) sources are a necessary
ingredient for phenomenological recipes aimed at describing the
spectrum of cosmological background at different wavelengths
\cite{Comastri}, as well as the physics and ionization properties of
the IGM \cite{Haardt}. Programs of multi-wavelength imaging and
spectroscopic follow-up of the distant $X$--ray sources identified by
Chandra and XMM--Newton are currently underway
\cite{Tozzi,Hasinger,Willott,Gandhi,Miyaji}. Once completed, they will
provide invaluable information about the process of galaxy formation
and its interplay with the history of cosmic baryons.


\section*{Acknowledgments} It is a pleasure to thank the Conference
Organizers, particularly Doris Neumann, for setting up such a
successful meeting in the wonderful environment of Les Arcs.

\end{document}